\documentstyle[12pt]{article}
\begin{document}

\begin{flushright}{UT-910\\ September 2000}
\end{flushright}
\vskip 0.5 truecm
%\vskip 0.5 truecm

\begin{center}
{\large{\bf Quantum and Classical Gauge Symmetries\footnote{Talk
given at ICHEP2000, Osaka, Japan, July 2000 (to be published in 
the Proceedings (World Scientific, Singapore))}}}
\end{center}
\vskip .5 truecm
\centerline{\bf Kazuo Fujikawa and Hiroaki Terashima}
\vskip .4 truecm
\centerline {\it Department of Physics,University of Tokyo}
\centerline {\it Bunkyo-ku,Tokyo 113,Japan}
\vskip 0.5 truecm

\begin{abstract}
The use of the mass term of the gauge field as a gauge fixing
term, which was discussed by Zwanziger, Parrinello and 
Jona-Lasinio in a large mass limit, is related to the 
non-linear gauge by Dirac and Nambu. 
We have recently shown that this use of the mass term as a 
gauge fixing term is in fact identical to the conventional 
 local Faddeev-Popov formula without taking a large mass 
limit, if one takes into account the variation of the gauge field 
along the entire gauge orbit. 
This suggests that the classical massive vector theory, 
for example, could be re-interpreted as a gauge invariant 
theory with a gauge fixing term added in suitably quantized 
theory.  
 As for massive gauge particles, the Higgs 
mechanics, where the  mass term is gauge invariant, has a more 
intrinsic meaning. We comment on  several implications of 
this observation.

\end{abstract}

\section{Introduction}
The Faddeev-Popov formula\cite{faddeev} and the resulting BRST
symmetry\cite{brs} provide a basis for the modern quantization
of gauge theory. On the other hand, a modified
quantization scheme\cite{zwanziger}\cite{jona-lasinio}
\begin{equation}
\int{\cal D}A_{\mu}\{\exp[-S_{YM}(A_{\mu})-
\int f(A_{\mu})dx]/\int{\cal 
D}g\exp[-\int f(A_{\mu}^{g})dx]\}
\end{equation}
with, for example, $f(A_{\mu})=(m^{2}/2)(A_{\mu})^{2}$
has been recently analyzed in a large mass limit in connection 
with the analysis of Gribov-type complications\cite{gribov}.
This gauge fixing in the large mass limit is related to the limit
$\lambda=0$ of the non-linear gauge 
\begin{equation}
A_{\mu}^{2}=\lambda =\ \ const.
\end{equation}
discussed by Dirac and Nambu\cite{nambu} many years ago. 
 Nambu used the above gauge to analyze the 
possible spontaneous breakdown of Lorentz symmetry. In his
treatment, the limit $\lambda=0$ is singular, and thus the 
present formulation is not quite convenient for an analysis of 
the possible breakdown of Lorentz symmetry.
Some aspects of this non-linear gauge have been discussd in 
Ref[7]. The above gauge fixing (1) has also been used 
in lattice simulation\cite{golterman}.
 
We have recently shown\cite{fujikawa} that the above scheme (1) 
is in fact identical  to the 
conventional {\em local} Faddeev-Popov formula
\begin{equation}
\int{\cal D}A_{\mu}\delta(D^{\mu}\frac{\delta f(A_{\nu})}
{\delta A_{\mu}})\det\{\delta[D^{\mu}
\frac{\delta f(A_{\nu}^{g})}{\delta A_{\mu}^{g}}]/\delta g\}
\exp[-S_{YM}(A_{\mu})]
\end{equation}
{\em without} taking the large mass limit, if one takes into 
account the variation of the gauge field along the entire gauge 
orbit parametrized by the gauge parameter $g$. 
The above  equivalence is valid only if the Gribov-type 
complications are ignored.

We here comment on  the possible implications\cite{fujikawa2} of 
the above equivalence, which is established without
taking the large mass limit, in a more general context of 
quantum gauge symmetry, namely, BRST symmetry.

\section{Abelian example}

We first briefly illustrate the proof\cite{fujikawa} of  the 
above equivalence of (1) and (3) by using an 
example of  Abelian gauge theory, $S_{0}=-(1/4)\int dx
(\partial_{\mu}A_{\nu}-\partial_{\nu}A_{\mu})^{2}$, 
for which we can work out 
everything explicitly. In this note we exclusively work on  
Euclidean theory with metric convention $g_{\mu\nu}=(1,1,1,1)$. 
As a simple and useful example, we choose
the gauge fixing function
$f(A)\equiv (1/2)A_{\mu}A_{\mu}$
and 
\begin{equation}
D_{\mu}(\frac{\delta f}{\delta A_{\mu}})=\partial_{\mu}A_{\mu}.
\end{equation}
Our claim above  suggests the relation
\begin{eqnarray}
Z&=&\int {\cal D}A^{\omega}_{\mu}\{e^{-S_{0}(A^{\omega}_{\mu})-
\int dx \frac{1}{2}(A^{\omega}_{\mu})^{2}}/\int {\cal D}h 
e^{-\int dx 
\frac{1}{2}(A^{h\omega}_{\mu})^{2}}\}\nonumber\\
&=&\int {\cal D}A^{\omega}_{\mu}{\cal D}B{\cal D}\bar{c}{\cal D}c 
e^{-S_{0}(A^{\omega}_{\mu})+ \int[-iB\partial_{\mu}
A^{\omega}_{\mu}+ 
\bar{c}(-\partial_{\mu}\partial_{\mu})c]dx }
\end{eqnarray}
where the variable $A^{\omega}_{\mu}$ stands for the field
variable obtained from $A_{\mu}$ by a gauge transformation
parametrized by the gauge orbit parameter $\omega$. 
To establish this result, we first evaluate 
\begin{eqnarray}
&&\int {\cal D}h e^{-\int dx \frac{1}{2}(A^{h\omega}_{\mu})^{2}} 
\nonumber\\
&&=\int {\cal D}h e^{-\int dx \frac{1}{2}(A^{\omega}_{\mu}
+\partial_{\mu}
h)^{2}} \nonumber\\
&&=\int {\cal D}h e^{-\int dx 
\frac{1}{2}[(A^{\omega}_{\mu})^{2}-2(\partial_{\mu}
A^{\omega}_{\mu})h + 
h(-\partial_{\mu}\partial_{\mu})h]} 
\nonumber\\
&&=\int {\cal D}B\frac{1}{det\sqrt{-\partial_{\mu}
\partial_{\mu}}} 
e^{-\int dx \frac{1}{2}[(A^{\omega}_{\mu})^{2}
-2(\partial_{\mu}A^{\omega}
_{\mu})\frac{1}{\sqrt{-\partial_{\mu}\partial_{\mu}}}B + 
B^{2}]}\nonumber\\
&&=\frac{1}{det \sqrt{-\partial_{\mu}\partial_{\mu}}}
e^{-\int dx \frac{1}
{2}(A^{\omega}_{\mu})^{2}+\frac{1}{2}\int\partial_{\mu}
A^{\omega}_{\mu}\frac{1}{-\partial_{\mu}\partial_{\mu}}
\partial_{\nu}A^{\omega}_{\nu}dx} 
\end{eqnarray}
where we defined $\sqrt{-\partial_{\mu}\partial_{\mu}}h=B$.
 Thus
\begin{eqnarray}
Z&=&\int{\cal D}A^{\omega}_{\mu}\{det 
\sqrt{-\partial_{\mu}\partial_{\mu}}\}e^{-S_{0}(A^{\omega}_{\mu})
-\frac{1
}{2}\int\partial_{\mu}A^{\omega}_{\mu}\frac{1}{-\partial_{\mu}
\partial_{\mu}}
\partial_{\nu}A^{\omega}_{\nu}dx}
\nonumber\\
&=&\int{\cal D}A^{\omega}_{\mu}{\cal D}B{\cal D}\bar{c}{\cal D}c
e^{-S_{0}(A^{\omega}_{\mu})-\frac{1}{2}\int B^{2}dx +\int
[-iB\frac{1}{\sqrt{-\partial_{\mu}\partial_{\mu}}}
\partial_{\mu}A^{\omega
}_{\mu}+\bar{c}\sqrt{-\partial_{\mu}\partial_{\mu}}c]dx}
\end{eqnarray}
which was derived in Refs.[3] and [4] and is invariant under the 
BRST transformation
\begin{eqnarray}
&&\delta A_{\mu}^{\omega}=i\lambda\partial_{\mu}c,\ \ \ 
\delta c=0\nonumber\\
&&\delta\bar{c}=\lambda B, \ \ \ 
\delta B=0
\end{eqnarray}
with a Grassmann parameter $\lambda$. Note the appearance of the
imaginary factor $i$ in the term
$iB\frac{1}{\sqrt{-\partial_{\mu}\partial_{\mu}}}
\partial_{\mu}A^{\omega}
_{\mu}$ 
in (7).

We next rewrite the expression (7) as
\begin{eqnarray}
&&\int{\cal D}A^{\omega}_{\mu}{\cal D}B{\cal D}\Lambda{\cal 
D}\bar{c}{\cal D}c 
\delta(\frac{1}{\sqrt{-\partial_{\mu}\partial_{\mu}}}
\partial_{\mu}A^{\omega}_{\mu}-\Lambda)e^{-S_{0}(A^{\omega}_{\mu}
)
-\frac{1}{2}\int(B^{2}+2i\Lambda B)dx 
+\int\bar{c}\sqrt{-\partial_{\mu}\partial_{\mu}}cdx}\nonumber\\
&&=\int{\cal D}A^{\omega}_{\mu}{\cal D}\Lambda{\cal D}\bar{c}
{\cal D}c 
\delta(\frac{1}{\sqrt{-\partial_{\mu}\partial_{\mu}}}
\partial_{\mu}
A^{\omega}_{\mu}-\Lambda)e^{-S_{0}(A^{\omega}_{\mu})-\frac{1}{2}
\int\Lambda^{2}dx +\int\bar{c}\sqrt{-\partial_{\mu}
\partial_{\mu}}cdx}.
\end{eqnarray}
We note that we can compensate any variation of 
$\delta\Lambda$ by a suitable change of gauge parameter 
$\delta\omega$ inside the $\delta$-function as 
\begin{equation}
\frac{1}{\sqrt{-\partial_{\mu}\partial_{\mu}}}\partial_{\mu}
\partial_{\mu
}\delta\omega=\delta\Lambda.
\end{equation}
By a repeated application of infinitesimal gauge transformations 
combined with the invariance of the path integral measure under 
these gauge transformations, we can re-write the formula (9) 
as 
\begin{eqnarray}
&&\int{\cal D}A^{\omega}_{\mu}{\cal D}\Lambda{\cal D}\bar{c}
{\cal D}c 
\delta(\frac{1}{\sqrt{-\partial_{\mu}\partial_{\mu}}}
\partial_{\mu}
A^{\omega}_{\mu})e^{-S_{0}(A^{\omega}_{\mu})-\frac{1}{2}
\int\Lambda^{2}dx 
+\int\bar{c}\sqrt{-\partial_{\mu}\partial_{\mu}}cdx}\nonumber\\
&&=\int{\cal D}A^{\omega}_{\mu}{\cal D}\bar{c}{\cal D}c 
\delta(\frac{1}{\sqrt{-\partial_{\mu}\partial_{\mu}}}
\partial_{\mu}
A^{\omega}_{\mu})e^{-S_{0}(A^{\omega}_{\mu})+\int\bar{c}
\sqrt{-\partial_{\mu}\partial_{\mu}}cdx}\nonumber\\
&&=\int{\cal D}A^{\omega}_{\mu}{\cal D}B{\cal D}\bar{c}{\cal D}c 
e^{-S_{0}(A^{\omega}_{\mu})+\int[-iB\frac{1}{\sqrt{-\partial_{\mu
}
\partial_{\mu}}}\partial_{\mu}A^{\omega}_{\mu}+\bar{c}
\sqrt{-\partial_{\mu}\partial_{\mu}}c]dx}\nonumber\\
&&=\int{\cal D}A^{\omega}_{\mu}{\cal D}B{\cal D}\bar{c}{\cal D}c 
e^{-S_{0}(A^{\omega}_{\mu})+\int[-iB\partial_{\mu}
A^{\omega}_{\mu}+\bar{c
}(-\partial_{\mu}\partial_{\mu})c]dx}.
\end{eqnarray}
In the last stage of this equation, we re-defined the 
{\em auxiliary} variables $B$ and $\bar{c}$ as 
\begin{equation}
B\rightarrow B\sqrt{-\partial_{\mu}\partial_{\mu}}, \ \ \ 
\bar{c}\rightarrow \bar{c}\sqrt{-\partial_{\mu}\partial_{\mu}} 
\end{equation}
which is consistent with BRST symmetry and leaves the path 
integral measure invariant. We have thus established the desired 
result (5). We emphasize that the integral over the entire 
gauge orbit, as is indicated in (10), is essential to derive a 
local theory (11) without taking a large mass
limit\cite{fujikawa}. 

It is shown that this procedure works for the non-Abelian case 
also\cite{fujikawa}, though the actual procedure is much more 
involved, if the (ill-understood) Gribov-type 
complications can be ignored such as in perturbative 
calculations.

\section{Possible Implications}

In the classical level, we traditionally consider
\begin{equation}
{\cal L}= -\frac{1}{4}
(\partial_{\mu}A_{\nu}-\partial_{\nu}A_{\mu})^{2}-\frac{1}{2}
m^{2}A_{\mu}A^{\mu}
\end{equation}
as a Lagrangian for a massive vector theory, and 
\begin{equation}
{\cal L}_{eff}= -\frac{1}{4}
(\partial_{\mu}A_{\nu}-\partial_{\nu}A_{\mu})^{2}-\frac{1}{2}
(\partial_{\mu}A^{\mu})^{2}
\end{equation}
as an effective Lagrangian for Maxwell theory with a Feynman-type 
gauge fixing term added. The physical meanings of these two 
Lagrangians are thus completely different. 

However, the analysis in Section 2 shows that the Lagrangian 
(13) could in fact be interpreted as a gauge fixed Lagrangian 
of {\em massless} Maxwell field in quantized theory.  To be 
explicit, by using (5), the Lagrangian (13) may be regarded as
 an effective Lagrangian in 
\begin{eqnarray}
Z&=&\int {\cal D}A^{\omega}_{\mu}\{e^{\int dx[ -\frac{1}{4}
(\partial_{\mu}A_{\nu}-\partial_{\nu}A_{\mu})^{2}-\frac{1}{2}
m^{2}A_{\mu}^{\omega}A^{\omega\mu}]}
/\int {\cal D}h e^{-\int dx
\frac{m^{2}}{2}(A^{h\omega}_{\mu})^{2}}\}\nonumber\\
&=&\int{\cal D}A^{\omega}_{\mu}{\cal D}B{\cal D}\bar{c}{\cal D}c 
e^{\int dx [-\frac{1}{4}(\partial_{\mu}A_{\nu}-
\partial_{\nu}A_{\mu})^{2}-iB\partial_{\mu}A^{\omega}_{\mu}+ 
\bar{c}(-\partial_{\mu}\partial_{\mu})c]}.
\end{eqnarray}
where we absorbed the factor $m^{2}$ into the definition of 
$B$ and $\bar{c}$.

One can also analyze (14) by defining 
$f(A_{\mu})\equiv\frac{1}{2}(\partial_{\mu}A^{\mu})^{2}$
in the modified quantization scheme (1). The equality of 
(1) and (3) then gives
\begin{eqnarray}
&&\int{\cal D}A_{\mu}\delta(D^{\mu}\frac{\delta f(A_{\nu})}
{\delta A_{\mu}})\det\{\delta[D^{\mu}
\frac{\delta f(A_{\nu}^{g})}{\delta A_{\mu}^{g}}]/\delta g\}
\exp[-S_{0}(A_{\mu})]\nonumber\\
&&=\int{\cal D}A_{\mu}\delta(\partial_{\nu}\partial^{\nu}
(\partial^{\mu}A_{\mu}))\det[\partial_{\nu}\partial^{\nu}
\partial_{\mu}\partial^{\mu}]
\exp[-S_{0}(A_{\mu})]\\
&&=\int{\cal D}A_{\mu}{\cal D}B{\cal D}\bar{c}{\cal D}c
\exp\{-S_{0}(A_{\mu})
+\int dx[-iB\partial_{\nu}\partial^{\nu}(\partial^{\mu}A_{\mu})
-\bar{c}(\partial_{\nu}\partial^{\nu}
\partial_{\mu}\partial^{\mu})c]\}\nonumber
\end{eqnarray}
After the re-definition of {\em auxiliary}
variables, $
B\partial_{\nu}\partial^{\nu} \rightarrow B,\ \ \ \
\bar{c}\partial_{\nu}\partial^{\nu} \rightarrow \bar{c}
$, which preserves BRST symmetry, (16) becomes
\begin{equation}
\int{\cal D}A_{\mu}{\cal D}B{\cal D}\bar{c}{\cal D}c
\exp\{-S_{0}(A_{\mu})
+\int dx[-iB(\partial^{\mu}A_{\mu})
+\bar{c}(-\partial_{\mu}\partial^{\mu})c]\}
\end{equation}
which agrees with (11) and (15).
We can thus assign an identical physical meaning to two 
classical Lagrangians (13) and (14) in suitably {\em quantized}
 theory.\\

Similarly, the two classical Lagrangians related to  
Yang-Mills fields
\begin{equation}
{\cal L}= -\frac{1}{4}(\partial_{\mu}A_{\nu}^{a}-
\partial_{\nu}A_{\mu}^{a}+gf^{abc}A_{\mu}^{b}A_{\nu}^{c})^{2}
-\frac{m^{2}}{2}A_{\mu}^{a}A^{a\mu}
\end{equation}
and 
\begin{equation}
{\cal L}_{eff}= -\frac{1}{4}(\partial_{\mu}A_{\nu}^{a}-
\partial_{\nu}A_{\mu}^{a}+gf^{abc}A_{\mu}^{b}A_{\nu}^{c})^{2}
-\frac{1}{2}(\partial_{\mu}A^{a\mu})^{2}
\end{equation}
could be assigned an identical physical meaning as an effective 
gauge fixed Lagrangian associated with the quantum theory
defined by\cite{fujikawa}
\begin{equation}
\int{\cal D}A_{\mu}^{a}{\cal D}B^{a}{\cal D}\bar{c}^{a}{\cal D}
c^{a}
\exp\{-S_{YM}(A_{\mu}^{a})
+\int dx[-iB^{a}(\partial^{\mu}A_{\mu}^{a})
+\bar{c}^{a}(-\partial_{\mu}(D^{\mu}c)^{a}]\}
\end{equation}
which is invariant under BRST symmetry.
 
We have illustrated that the apparent ``massive gauge
field'' in the classical level has no intrinsic 
physical meaning. It can be interpreted either as a 
massive (non-gauge) vector theory, or as a gauge-fixed effective 
Lagrangian for a massless gauge field in quantized theory. 
In the framework of path integral, these different 
{\em interpretations} may also be  understood as a more flexible 
choice of the path integral measure than the classical Poisson 
bracket analysis suggests\cite{fujikawa2}: One 
choice of the measure 
\begin{eqnarray}
&&\int d\mu\exp\{\int dx[-\frac{1}{4}(\partial_{\mu}A_{\nu}^{a}-
\partial_{\nu}A_{\mu}^{a}+gf^{abc}A_{\mu}^{b}A_{\nu}^{c})^{2}
-\frac{m^{2}}{2}A_{\mu}^{a}A^{a\mu}]\}\nonumber\\
&&\equiv\int{\cal D}A_{\mu}\frac{1}{\int{\cal 
D}g\exp[-\int\frac{m^{2}}{2}(A^{a g}_{\mu})^{2}dx]\}}\\
&&\times\exp\{\int dx[-\frac{1}{4}(\partial_{\mu}A_{\nu}^{a}-
\partial_{\nu}A_{\mu}^{a}+gf^{abc}A_{\mu}^{b}A_{\nu}^{c})^{2}
-\frac{m^{2}}{2}A_{\mu}^{a}A^{a\mu}]\}\nonumber
\end{eqnarray}
gives rise to a renormalizable massless gauge theory, and the 
other naive choice
\begin{eqnarray}
&&\int d\mu\exp\{\int dx[-\frac{1}{4}(\partial_{\mu}A_{\nu}^{a}-
\partial_{\nu}A_{\mu}^{a}+gf^{abc}A_{\mu}^{b}A_{\nu}^{c})^{2}
-\frac{m^{2}}{2}A_{\mu}^{a}A^{a\mu}]\}\\
&&\equiv\int{\cal D}A_{\mu}
\exp\{\int dx[-\frac{1}{4}(\partial_{\mu}A_{\nu}^{a}-
\partial_{\nu}A_{\mu}^{a}+gf^{abc}A_{\mu}^{b}A_{\nu}^{c})^{2}
-\frac{m^{2}}{2}A_{\mu}^{a}A^{a\mu}]\}\nonumber
\end{eqnarray}
gives rise to a non-renormalizable massive 
{\em non-gauge} theory. A somewhat 
analogous situation arises when one attempts to quantize the 
so-called anomalous gauge theory: A suitable choice of the 
measure with a Wess-Zumino term gives rise to a consistent 
quantum theory in 2-dimensions, for example\cite{jackiw}.
From a view point of classical-quantum correspondence, one can 
define a classical theory uniquely starting from quantum theory 
by considering the limit $\hbar\rightarrow 0$, but not the other 
way around in general. 

In the context of the present general interpretation of 
apparently massive classical 
gauge fields, the massive gauge fields generated by the 
Higgs mechanism are exceptional and quite different. Since all
the terms including the mass term are gauge invariant, one
can assign an intrinsic meaning to the massive gauge field
in Higgs mechanism. In view of the well known fact that the 
massive non-Abelian gauge
theory is inconsistent as a quantum theory (22), it may be 
sensible to treat all the classical massive non-Abelian 
Lagrangians as a gauge fixed version of pure non-Abelian gauge
 theory and to restrict the massive non-Abelian gauge fields to 
those generated by the Higgs mechanism.

It is a long standing question if one can generate gauge 
fields from some {\em more} fundamental mechanism.
To our knowledge, however, there exists no definite convincing
scheme so far. On the contrary, there is a no-go theorem or 
several arguments against such an 
attempt\cite{case}. 
Apart from technical details, the basic argument against the 
``dynamical'' generation of gauge fields is that the  Lorentz
invariant positive definite theory cannot simply generate the 
negative metric states associated with the time components of 
massless gauge fields. In contrast, the dynamical 
generation of the Lagrangian of the structure 
\begin{equation}
{\cal L}=-\frac{1}{4}(\partial_{\mu}A_{\nu}^{a}-
\partial_{\nu}A_{\mu}^{a}+gf^{abc}A_{\mu}^{b}A_{\nu}^{c})^{2}
-\frac{m^{2}}{2}(A_{\mu}^{a})^{2}
\end{equation}
does not appear to be prohibited by general arguments so far.
If one considers that 
the induced Lagrangian such as (23) is a {\em classical} 
object which should be quantized anew, one could regard   
$\frac{m^{2}}{2}(A_{\mu}^{a})^{2}$, which breaks classical gauge 
symmetry, as a gauge fixing term in the modified 
quantization scheme\cite{zwanziger}\cite{jona-lasinio}. In this 
interpretation, one 
might be allowed to say that massless gauge fields are generated 
dynamically. Although a dynamical generation of pure gauge 
fields is prohibited, a {\em gauge fixed} Lagrangian might be 
allowed to be generated. (In this respect, one may recall that 
much of the arguments for the no-go 
theorem\cite{case} would be  refuted if one could 
generate a gauge fixed Lagrangian with the Faddeev-Popov term 
added.) The mass for the gauge field which has 
an intrinsic unambiguous physical meaning is then further 
induced by the spontaneous symmetry breaking of the gauge 
symmetry thus defined (the Higgs mechanism). 
 
We next comment on a mechanism for generating gauge fields by 
the violent random fluctuation of gauge degrees of freedom at 
the beginning of the universe\cite{nielsen}; this scheme
is based on the renormalization group flow starting from an 
initial chaotic theory. In such a scheme, it is natural to think 
that one is always dealing with quantum theory, and thus no room 
for our way of re-interpretation of the induced theory.
Nevertheless, we find a possible connection in the following
sense: To be precise, an example of massive Abelian gauge 
field in {\em compact} lattice gauge theory
\begin{equation}
\int{\cal D}U\frac{{\cal D}\Omega}{vol(\Omega)}
\exp[-S_{inv}(U)-S_{mass}(U^{\Omega})]
\end{equation}
is analyzed in Ref.[13].
Here $S_{inv}(U)$ stands for the gauge invariant part of 
the lattice Abelian gauge field $U$, and $S_{mass}(U^{\Omega})$
stands for the gauge non-invariant mass term with the gauge 
freedom $\Omega$. In compact theory, one need not fix the 
gauge and instead one may take an average over the entire gauge 
volume of $\Omega$.
They argued that the mass term, which breaks gauge symmetry
softly, disappears in the long distance limit when one integrates 
over the entire gauge freedom $\Omega$. Their scheme is 
apparently
dynamical one, in contrast to the kinematical nature of our 
re-interpretation. Nevertheless, the massive Abelian theory is a 
free theory in continuum formulation, and the disappearance of 
the mass term by a mere smearing over the gauge volume may 
suggest
 that the mass term in their scheme is also treated as a kind of
gauge artifact, just as in our kinematical re-interpretation. 
 
In conclusion, the equivalence of (1) and (3) allows a 
more flexible {\em quantum interpretation} of various classical 
Lagrangians such as massive gauge theory. 

  As for a recent BRST analysis of the observation in Ref.[10],
see Ref.[14].

\end{document}